# Theoretical models for NO decomposition in Cu-exchanged zeolites

Roumen Tsekov and P.G. Smirniotis
Chemical Engineering Department, University of Cincinnati, Cincinnati, Ohio 45221

A unified description of the catalytic effect of Cu-exchanged zeolites is proposed for the decomposition of NO. A general expression for the rate constant of NO decomposition is obtained by assuming that the rate-determining step consists of the transferring of a single atom associated with breaking of the N-O bond. The analysis is performed on the base of the generalized Langevin equation and takes into account both the potential interactions in the system and the memory effects due to the zeolite vibrations. Two different mechanisms corresponding to monomolecular and bimolecular NO decomposition are discussed. The catalytic effect in the monomolecular mechanism is related to both the $Cu^+$ ions and zeolite O-vacancies, while in the case of the bimolecular mechanism the zeolite contributes through dissipation only. The comparison of the theoretically calculated rate constants with experimental results reveals additional information about the geometric and energetic characteristics of the active centers and confirms the logic of the proposed models.

Nitric oxides emitted by both mobile and stationary sources cause serious environmental problems. The demand of methods to reduce NO pollution has initiated intensive studies on NO degradation chemistry[1]. The most desirable process is to directly decompose the NO molecule to its elements, $N_2$ and $O_2$. This process is thermodynamically favorable, but its rate is very low. Thus, an important problem is to find active catalysts to substantially accelerate the reaction rate of NO decomposition. Several tests with well-known metal catalysts have demonstrated that they inefficiently catalyze NO decomposition; the catalysts are oxidized when the reaction progresses and finally the NO decomposition is inhibited. The most active catalysts for NO decomposition are ion-exchanged zeolites and primarily Cu-ZSM-5. These catalysts are prepared by introducing metal ions in zeolites via standard ion-exchange procedures[2]. Despite the large number of publications after the initial report of the catalytic activity of Cu-ZSM-5, there is still ongoing discussion about the catalytic mechanism. Iwamoto and Hamada[3], as well as Hall and Valyon[4], suggested that the active sites were $Cu^+$ ions rather than $Cu^{2+}$. Hence, the copper redox chemistry[5] is of crucial importance for the catalytic performance. Selef[6], however, has proposed that the square planar complex $Cu^{2+}(NO)_2$ is an alternative intermediate state responsible for the reaction. Kuroda et al.[7] have discovered an effect of O-bridged copper planar complexes while Shpiro et al.[8] have proposed that CuO clusters are the active catalytic centers. There is also no clear trend on the catalytic effect of the Si/Al ratio of zeolites. Moretti[9] reported that an increase of the Si/Al ratio leads to a decrease in catalytic activity thus indicating that the active centers include two or more Cu-substituted basic centers. On the contrary, Wichterlova et al.[10] demonstrated that there is a correlation between the catalytic activity and the number of Cu-ions attached on a single basic center.

Numerous modern experimental methods were employed in the investigations of NO decomposition such as isotopic labeling[11] and IR spectroscopy[12]. Advanced theoretical models based on first principal calculations[13], statistical mechanics[14], computer modelling[15] and simulations[16] have also been applied[17]. Identification of the active sites for NO decomposition can probably be essential for the selective catalytic reduction of NO. It is well known that in the selective catalytic reduction of NO by hydrocarbons, the important step is the adsorption-excitation of the NO molecule. Hence, the NO SCR mechanism is related[18] to that of NO decomposition. The present theoretical study aims at the development of a unified theory for the reaction rate of NO decomposition over Cu-exchanged zeolites. The model is based on the presumption that the rate-determining step is a single atom transfer. The motion of the transferred atom is described by a generalized Langevin equation accounting for both the interactions of the transferred atom with the rest of the NO molecule and the zeolite framework. Thus, the memory effects of the friction force are accounted for, which reflect the spectrum of vibrations in the system. Using the Grote-Hynes theory, the rate constant of this chemical transformation is calculated. The proposed model was examined for two possible reaction schemes for NO decomposition describing monomolecular and bimolecular mechanisms.

There are many processes involved in the chemical transformations of adsorbed species[19], such as electron transfer between an adsorbed molecule and the adsorbent, molecule polarization, vibration energy exchange, etc. A useful concept in chemical kinetics is that the rate of chemical transformation is equal to the rate of the slowest step of the reaction scheme. Hence, instead of describing the whole process, it is only necessary to look for the rate-determining step. The decomposition of NO over Cu-zeolites is relatively easy to treat because of the relatively simple nature of the NO molecule. Obviously, this process involves dissociation of the nitrogen oxide molecule. The basic assumption of the present theory is that the electronic processes are fast, and the rate-determining step is the rearrangement of atoms. Moreover, the NO molecule is adsorbed on the zeolite surface, and both the N and O atoms are influenced by the interaction with the zeolite. Hence, the breaking of the chemical bond can be simplified to a single atom transfer leading to the rupture of the initial N-O bond and creation of a new one.

Let us consider the simplest case where the reaction coordinate between the initial and the final state of the transferred atom is a straight line. Since the chemical transformations are carried out at relatively high temperatures, the quantum effects are negligible and the motion of the atoms of the NO molecule and zeolite can be satisfactorily described in the frames of classical mechanics. Assuming a linear coupling between the zeolite vibrations and the transferred atom, the equations of motion acquire the forms[20]

$$M\ddot{X} = -\partial_X \Phi_0 - \sum_q (\partial_X \partial_q \Phi)_0 q \qquad (1)$$

$$\ddot{q} + \omega_q^2 q = -(\partial_q \Phi)_0 - (\partial_q^2 \Phi)_0 q \qquad (2)$$

Here $M$ is the mass of the transferred atom, $X$ is its coordinate along the reaction path, and $\Phi$ is the interaction potential of the transferred atom and the rest of the atoms in the system. Since the latter atoms are considered only to vibrate along their equilibrium positions, their motion is described with a set of normal modes and frequencies $\{q, \omega_q\}$. For the sake of simplicity, the coupling of the normal modes due to the interaction potential $\Phi$ is neglected in eq 2. The subscript $_0$ in the equations above refers to $q = 0$. Hence, $\Phi_0$ is the potential energy of the transferred atom resulting from the interaction with other atoms fixed at their equilibrium positions. Solving eq 2 and introducing the result for $q(t)$ into eq 1 one can obtain the following generalized Langevin equation

$$M\ddot{X} + \int_0^t ds\, G(t-s)\dot{X}(s) + \partial_x U = F \qquad (3)$$

$$G(\tau) = \sum_q \frac{(\partial_X \partial_q \Phi)_0^2}{\omega_q^2 + (\partial_q^2 \Phi)_0} \cos[\sqrt{\omega_q^2 + (\partial_q^2 \Phi)_0}\ \tau] \qquad (4)$$

$$U = \Phi_0 - \frac{1}{2}\sum_q \frac{(\partial_q \Phi)_0^2}{\omega_q^2 + (\partial_q^2 \Phi)_0} \qquad (5)$$

which is applicable to the diffusion in a potential force field[21]. The kernel $G$ is called the memory function, since it accounts for the effect of past motion of the atom in its contemporary trajectory. The effective potential $U$ represents both the static potential $\Phi_0$ and a term due to the average displacement of the atoms from their equilibrium positions caused by the transferred atom. Finally, the Langevin force $F$ is a stochastic one with zero mean value and autocorrelation $<F(t)F(t+\tau)>= k_B T G(\tau)$ where $T$ is the temperature and $k_B$ is the Boltzmann constant.

Equation 3 describes the dynamics of atomic transfer from the initial to the final position under the action of the surroundings. In general, there are three types of interaction between the atom and environment namely the friction (the corresponding force is given by the integral in eq 3), the potential force of the effective field $U$, and the random fluctuation force $F$. The fact that the potential $\Phi$ approaches zero rapidly as the distance between the transferred and other atoms increases is a convenient circumstance[20]. It allows all the terms in eqs 4 and 5 to be expressed by the static potential $\Phi_0$. In this way the results above can be simplified to

$$G = \frac{(\partial_X^2 \Phi_0)^2}{M} \Gamma(\tau) \qquad (6)$$

$$U = \Phi_0 - \frac{\Gamma(0)}{2M}(\partial_X \Phi_0)^2 \qquad (7)$$

$$\Gamma(\tau) = \int_0^\infty \frac{\cos(\tau\sqrt{\omega^2 + \partial_X^2 \Phi_0 / M})}{\omega^2 + \partial_X^2 \Phi_0 / M} g(\omega) d\omega \qquad (8)$$

where $\Gamma$ is the time dependent memory function and $g$ is the spectral density of all the vibrations in the system, excluding the transferred atom. Note that $g$ is strongly influenced by the close surrounding of the transferred atom and hence it accounts for the $Cu^+$ ion and the adsorbed atom of the NO molecule in addition to the zeolite vibrations.

In general, the initial and final state of the transferred atom corresponds to minima of the potential $U$, while the transition state is defined at the $U$-maximum. Since the last term in eq 7 is proportional to $(\partial_X \Phi_0)^2$, the minima and maxima of the potential $U$ coincide with those of the potential $\Phi_0$. Hence, the activation energy to escape over the potential barrier is not affected by the lattice displacement from the equilibrium position. It is well known that the rate constant of the considered atom transfer satisfies the Arrhenius law

$$k = k_0 \exp(-\frac{\varepsilon_a}{k_B T}) \qquad (9)$$

where $k_0$ is a temperature independent pre-exponential factor and $\varepsilon_a$ is the activation energy being the difference of the potential energies of the transition and initial states. If we denote by $b$ the coordinate of the atom on the top of the barrier and by $a$ its coordinate at the bottom of the initial potential well $[(\partial_X U)_a = (\partial_X \Phi_0)_a = (\partial_X U)_b = (\partial_X \Phi_0)_b = 0]$, the activation energy is equal to

$$\varepsilon_a = U(b) - U(a) = \Phi_0(b) - \Phi_0(a) \qquad (10)$$

According to the Grote-Hynes theory[22] the pre-exponential factor $k_0$ is equal to

$$k_0 = \frac{z\omega_a}{2\pi\omega_b} \tag{11}$$

where $\omega_a$ is the frequency of the vibrations around the initial state, which is defined via the relation $M\omega_a^2 \equiv (\partial_X^2 U)_a = M\Omega_a^2[1-\Gamma_a(0)\Omega_a^2]$ with $M\Omega_a^2 = (\partial_X^2 \Phi_0)_a$. The frequency $\omega_b$ is associated with the curvature of the potential on the top of the barrier and can be calculated from $M\omega_b^2 \equiv -(\partial_X^2 U)_b = M\Omega_b^2[1+\Gamma_b(0)\Omega_b^2]$ with $M\Omega_b^2 = -(\partial_X^2 \Phi_0)_b$. Finally, the frequency $z$ is the solution of the following equation

$$z^2 + \Omega_b^4 \tilde{\Gamma}_b(z) z = \omega_b^2 \tag{12}$$

Here $\tilde{\Gamma}_b$ is the Laplace transformation of the time-dependent memory function

$$\tilde{\Gamma}_b(z) = \int_0^\infty \Gamma_b(\tau) \exp(-z\tau) d\tau = \int_0^\infty \frac{z}{(\omega^2 - \Omega_b^2)(z^2 + \omega^2 - \Omega_b^2)} g(\omega) d\omega \tag{13}$$

As seen knowledge of the potential $\Phi_0$ allows the calculation of the activation energy and the pre-exponential factor, and the only additional information needed is the spectral density of the vibrations $g$.

The simplest model of zeolite vibrations is the Debye one $3\omega^2/\omega_D^3 \; H(\omega_D - \omega)$, where $H$ is the Heaviside step function and $\omega_D$ is the Debye frequency. However, since the spectral density $g$ also includes the vibrations of the adsorbed species, their characteristic spectra should be accounted for, too. Usually, there is a dominant frequency. For this reason we are going to describe the effect of the adsorbed species by the Einstein model $2\delta(\omega_E - \omega)$, where the Einstein frequency $\omega_E$ represents the dominant frequency of the species vibrations. Hence, a realistic model of the spectral density of the vibrations in the system is a linear combination between the Debye and Einstein models

$$g(\omega) = (1-\alpha)2\delta(\omega_E - \omega) + \alpha \frac{3\omega^2}{\omega_D^3} H(\omega_D - \omega) \tag{14}$$

where the parameter $\alpha$ is a number between 0 and 1. Since the transferred atom interacts strongly with the atoms of the adsorbed species (NO, Cu), it is reasonable to expect the parameter $\alpha$ to be a small number. In addition, our previous calculations[23] show that there is no sub-stantial difference in the rate constants calculated by using the Debye and Einstein models, respectively. Hence, we will assume that $\alpha = 0$. Manipulating eq 14 in this way allows one to calculate the Laplace image of the time-dependent correlation function given by eq 13, which results in

$$\tilde{\Gamma}_b(z) = \frac{z}{(\omega_E^2 - \Omega_b^2)(z^2 + \omega_E^2 - \Omega_b^2)} \tag{15}$$

The corresponding zero time values are given by

$$\Gamma_a(0) = \int_0^\infty \frac{g(\omega)}{\omega^2 + \Omega_a^2} d\omega = \frac{1}{\omega_E^2 + \Omega_a^2} \qquad \Gamma_b(0) = \int_0^\infty \frac{g(\omega)}{\omega^2 - \Omega_b^2} d\omega = \frac{1}{\omega_E^2 - \Omega_b^2} \tag{16}$$

By using eq 16, it is easy to calculate the characteristic frequencies at the initial state and at the top of the barrier, respectively

$$\omega_a = \frac{\omega_E \Omega_a}{\sqrt{\omega_E^2 + \Omega_a^2}} \qquad \omega_b = \frac{\omega_E \Omega_b}{\sqrt{\omega_E^2 - \Omega_b^2}} \tag{17}$$

Finally, by introducing expression 15 into eq 12 and solving the resulting biquadratic equation, the parameter $z$ equals

$$z = \sqrt{\frac{\sqrt{\omega_E^4 + 4\Omega_b^4} - \omega_E^2 + 2\Omega_b^2}{2}}$$

Substituting this expression for $z$ and eq 17 into eq 11, the pre-exponential factor of the rate constant acquires the form

$$k_0 = \frac{\Omega_a}{2\pi \Omega_b} \sqrt{\frac{(\sqrt{\omega_E^4 + 4\Omega_b^4} - \omega_E^2 + 2\Omega_b^2)(\omega_E^2 - \Omega_b^2)}{2(\omega_E^2 + \Omega_a^2)}} \tag{18}$$

The activation energy can be calculated from eq 10, while the pre-exponential factor can be obtained from eq 18 after calculation of the characteristic frequencies $\Omega_a$ and $\Omega_b$ of the potential $\Phi_0$. Hence, in order to calculate the rate constant of NO decomposition from eq 9, knowledge of the static interaction potential $\Phi_0$ and the Einstein frequency $\omega_E$ is required. The theory described above will be applied for two reaction models corresponding to monomolecular and bimolecular mechanisms.

The monomolecular mechanism assumes that the rate determining step during the NO decomposition in zeolites is the activation of an adsorbed NO molecule. A slightly modified scheme of the mechanism proposed by Li and Hall[24] is presented in Figure 1. The first step is the desorption of oxygen from the zeolite lattice close to an adsorbed $Cu^{2+}$ ion. As a result of this, the $Cu^{2+}$ ion reduces to $Cu^+$, and a localized positively charged O-vacancy is created in the zeolite framework. This structure involving $Cu^+$ and the O-vacancy is the active site for NO decomposition. After the adsorption of a NO molecule on this site, $Cu^+$ restores to its initial state $Cu^{2+}$; this has been observed in many experiments. Thus the double bond of the NO molecule reduces to a single bond, which is the first catalytic effect of the copper ion. The second catalytic effect is the electrostatic attraction between the negatively charged oxygen atom of the NO molecule and the positively charged zeolite oxygen vacancy. Then, the transfer of the oxygen atom from the NO molecule to the zeolite takes place, which is the rate-determining step. The desorption of the nitrogen atom and recovery of the initial state follows.

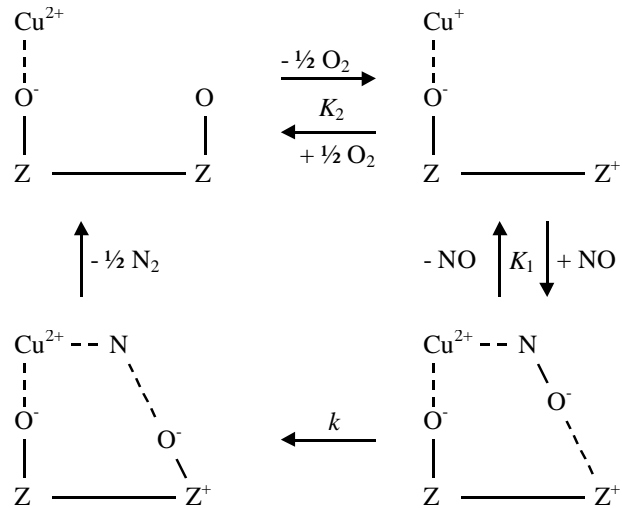

**Fig. 1** Scheme of the monomolecular mechanism of NO decomposition
Over Cu-exchanged zeolites

The turnover frequency *TOF* is defined by the number of NO molecules decomposed per unit time per $Cu^{2+}$ ion. Since the rate-determining step is the dissociation of the NO molecule, the turnover frequency is equal to

$$TOF = k\frac{[Cu^{2+}NO^-]}{[Cu^{2+}]+[Cu^+]} = kK_1 p_{NO} \frac{[Cu^+]}{[Cu^{2+}]+[Cu^+]} = kK_1 K_2 \frac{p_{NO}}{p_{O_2}^{1/2}} \frac{[Cu^{2+}]}{[Cu^{2+}]+[Cu^+]} \approx kK_1 K_2 \frac{p_{NO}}{p_{O_2}^{1/2}} \quad (19)$$

where $[Cu^{2+}NO^-]$, $[Cu^{2+}]$ and $[Cu^+]$ are the concentrations of $Cu^{2+}NO^-$ complexes, $Cu^{2+}$ and $Cu^+$ ions, respectively. The second equality follows from the adsorption-desorption equilibrium of NO molecules which is described with equilibrium constant $K_1$. The third equality implies adsorption-desorption equilibrium of oxygen with equilibrium constant $K_2$. Finally, since the number of active sites is relatively low as compared to the total number of copper ions, the ratio $[Cu^{2+}]/([Cu^{2+}]+[Cu^+])$ is close to unity[24], and therefore the turnover frequency relation can be simplified to the last expression in eq 19. The *TOF* is proportional to the pressure of nitrogen oxide and inversely proportional to the square root of oxygen pressure. A similar reaction rate was derived from the experiments of Li and Hall[24]. Calculation of the equilibrium constants $K_1$ and $K_2$ is beyond the scope of the present paper. Our theoretical model aims the determination of the reaction rate constant $k$.

According to Figure 1, the transferred O-atom interacts with the adsorbed N-atom and the oxygen vacancy in the zeolite framework. The N-O bonding can be described satisfactorily by a Morse potential, while the electrostatic attraction between the zeolite and the O-atom is subject to Coulomb's law. Hence, the static interaction potential acting on the transferred O-atom is

$$\Phi_0 = D\{\exp[2q(X_0 - X)] - 2\exp[q(X_0 - X)]\} - \frac{e^2}{4\pi\varepsilon_0(X_\infty - X)} \tag{20}$$

The values of the parameters in eq 20 are taken from the *CRC* chemistry handbook[25]; $D$ = 10$^{-18}$ J is the NO molecule bond strength, $X_0$ = 1.15 Å is the equilibrium bond length, and the parameter $q$ can be calculated from the N-O bond force constant $f = 2Dq^2$ = 15.95 N cm$^{-1}$. The electron charge and vacuum dielectric permittivity are equal to $e$ = 1.6 10$^{-19}$ C and $\varepsilon_0$ = 8.85 10$^{-12}$ F m$^{-1}$, respectively, while $X_\infty$ is the distance between the N-atom and the O-vacancy. Unfortunately, it is impossible to find any analytical expression for the coordinates *a* and *b* of the minimum and maximum of the potential given by eq 20. For this reason numerical calculations are necessary. In Figure 2 the dependence of the activation energy as a function of the distance of the O-vacancy from the N-atom is plotted. The increase of the distance between these atoms results in an increase of the activation energy which finally approaches a saturation value equal to $D$. A typical value[24] for the activation energy is $\varepsilon_a = 0.18D$, which according to Figure 2, corresponds to $X_\infty = 2.7X_0$, $a = 1.04X_0$ and $b = 1.42X_0$. The fact that the distance $X_\infty$ is only three times larger than $X_0$ supports the idea that the desorbed oxygen from the zeolite is from the very neighborhood of the adsorbed $Cu^{2+}$ ion. Since the minimum of the potential given by eq 20 is relatively far from the zeolite O-vacancy, the value of *a* is less affected by and is almost equal to $X_0$.

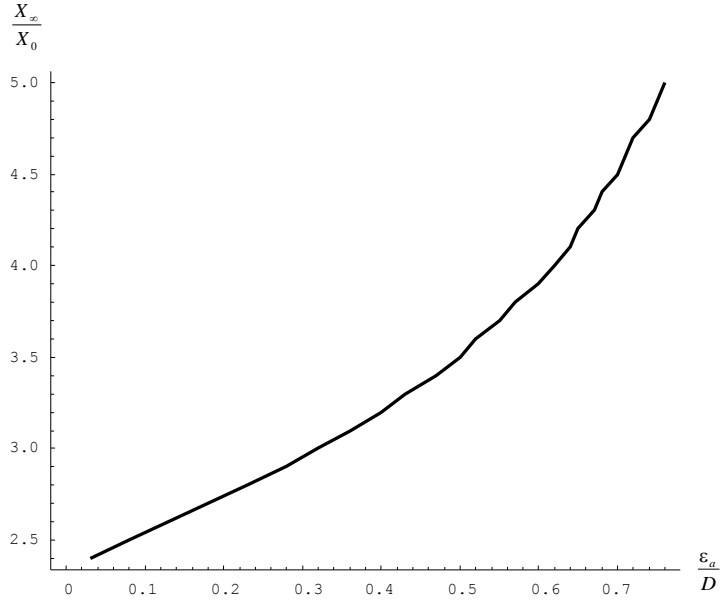

**Fig. 2** Dependence of the activation energy of NO decomposition from the distance between the nitrogen atom and the oxygen vacancy in the zeolite framework

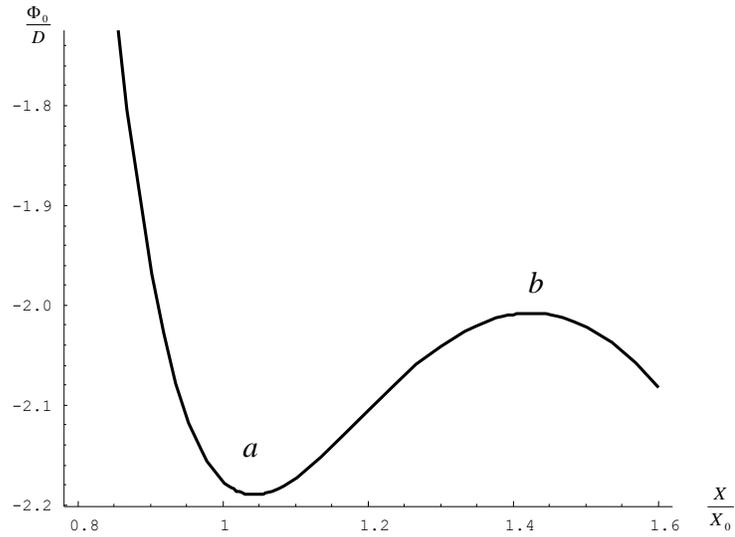

**Fig. 3** Potential energy of the oxygen atom bonding at $X_\infty = 2.7 X_0$: *a* corresponds to ZCu$^{2+}$NO$^-$ and *b* corresponds to Z(Cu$^{2+}$N)O$^-$.

The corresponding plot of the potential $\Phi_0$ for the value $X_\infty = 2.7 X_0$ is given in Figure 3. It is shown that there is a potential barrier that the oxygen atom of NO has to overcome in order to separate from nitrogen. Knowing the mass of an oxygen atom ($M$ = 16 g mol$^{-1}$), it is easy to numerically calculate the characteristic frequencies $\Omega_a = 187$ ps$^{-1}$ and $\Omega_b = 129$ ps$^{-1}$. Hence, in order to calculate the pre-exponential factor, the Einstein frequency is the only value needed. The dependence of $k_0$ versus $\omega_E$ is presented in Figure 4. The pre-exponential factor increases with increasing $\omega_E$ with a tendency to reach the transition state theory value of $\omega_a/2\pi$. This effect is due to the diminishing friction with increasing $\omega_E$. In the present mechanism, the most important frequency of the adsorbed species vibrations seems to be the N-O bond one at the potential well. Hence, the Einstein frequency can be estimated to be equal to $\Omega_a$; according to Figure 4 the pre-exponential factor is equal to $k_0 = 18$ ps$^{-1}$.

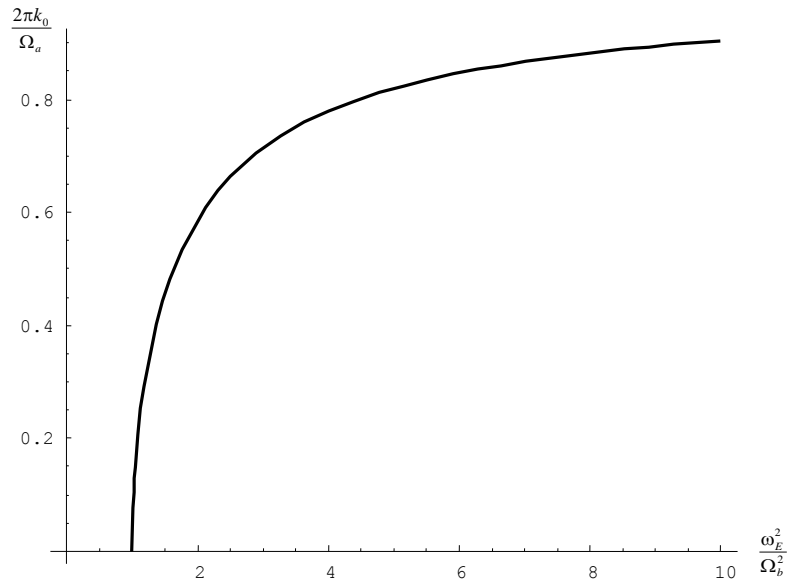

**Fig. 4** Dependence of the pre-exponential factor on the Einstein frequency of the local vibrations

Finally, let us demonstrate the applicability of the present theory for a real set of data. According to the experimental data of Li and Hall[24], $kK_1$ = 1.2 x 10$^{-6}$ Pa$^{-1}$s$^{-1}$ and $K_2$ = 30 Pa$^{1/2}$ at 773°K. On the other hand[14] the equilibrium constant of the NO adsorption at standard pressure and 773°K is $K_1 = 0.4$ Pa$^{-1}$. Hence, the rate constant is equal to $k$ = 3 x 10$^{-6}$ s$^{-1}$, which corresponds to an activation energy $\varepsilon_a$ = 0.46$D$ at $k_0 = 18$ ps$^{-1}$. Checking for this value in Figure 2, we see that it corresponds to $X_\infty = 3.4 X_0$. Thus, we have geometrically and energetically parameterized the active site for NO decomposition in the experiments of Li and Hall. The fact that the activation energy is about three times higher than the value considered before implies a substantial temperature dependence of the equilibrium constant $K_1$. Since Li and Hall have plotted Arrhenius plots of the product $kK_1$, they have calculated an apparent activation energy, which is equal to the difference between $\varepsilon_a$ and the latent head of NO adsorption.

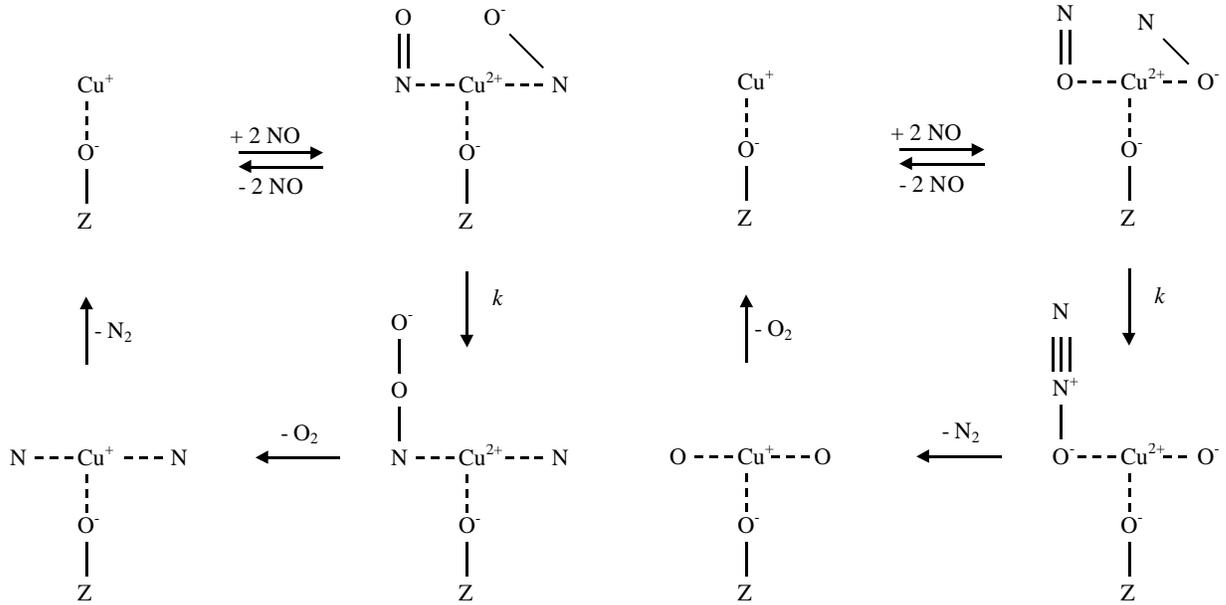

**Figs. 5 & 6** Scheme of bimolecular mechanism of NO decomposition over Cu-exchanged zeolites

The bimolecular mechanism, which is favored at relatively low temperatures, considers the adsorption of two NO molecules on a $Cu^+$ site[14]. Since the interactions between the copper ion and the N and O atoms of the NO molecule are commensurate the adsorption can occur in the two ways presented in Figures 5 and 6. In Figure 5 an oxygen atom is transferred to form the spectroscopically identified $NO_2^-$ intermediate[12]. In Figure 6 a nitrogen atom transfers and the intermediate formation of a $N_2O$ molecule is predicted by first-principle calculations[13]. It is a plausible explanation for the nitrous oxide co-product observed during NO decomposition.

For the sake of simplicity, we will assume that the interactions of the transferred atom with oxygen and nitrogen atoms are nearly the same. This assumption is supported by the data for the O-O, N-O and N-N bonds[25]. Hence, a unified potential for both the two schemes on Figures 5 and 6 is

$$\Phi_0 = D\{\exp[2q(X_0 - X)] - 2\exp[q(X_0 - X)]\} + D\{\exp[2q(X - X_\infty)] - 2\exp[q(X - X_\infty)]\} \quad (21)$$

This model describes the interaction of the transferred atom by two Morse potentials with $D = 10^{-18}$ J, $X_0 = 1.15$ Å and $qX_0 = 3.25$. Its plot at $X_\infty = 2X_0$ is presented in Figure 7. The minimum of the first Morse potential is at $X_0$, which corresponds to the initial bounded state of the transferred atom, while the minimum of the second Morse potential at $X_\infty$ corresponds to the final bounded state. The simplified potential $\Phi_0$ in eq. 21 allows one to calculate explicitly the coor-dinates of the initial and transition states, $a = X_0$ and $b = (X_\infty + X_0)/2$, respectively. If one introduces a dimensionless parameter $x = 13(1 - X_\infty / X_0)/8$, the activation energy and the two characteristic potential curvatures can be expressed analytically as follows

$$\varepsilon_a = D \ [1 - 4\exp(x) + 4\exp(2x) - \exp(4x)] \tag{22}$$

$$M\Omega_a^2 = \frac{13^2 D}{8 X_0^2}[1 - \exp(2x) + 2\exp(4x)] \tag{23}$$

$$M\Omega_b^2 = \frac{13^2 D}{4 X_0^2}[\exp(x) - 2\exp(2x)] \tag{24}$$

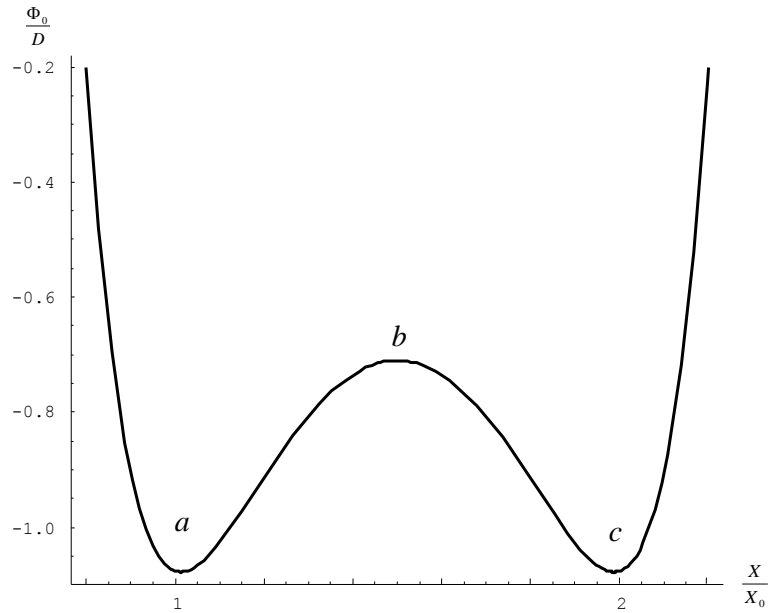

**Fig. 7** Potential energy of the transferred atom bonding at $X_\infty = 2X_0$: *a* corresponds to Z(Cu$^{2+}$NO$^-$)(NO) or ZCu$^+$(ON)$_2$, *b* corresponds to Z(Cu$^{2+}$N)(NO)O$^-$ or Z(Cu$^+$O)(ON)N, and *c* corresponds to Z(Cu$^{2+}$N)(NO$_2^-$) or Z(Cu$^+$O)(ON$_2$)

The dependencies of the activation energy $\varepsilon_a$ and the frequencies $\Omega_a$ and $\Omega_b$ on the ratio $X_\infty / X_0$ are plotted in Figures 8-10, respectively. With increasing distance between the final and initial bounded states, the activation energy increases to reach the value of the dissociation energy *D*. The frequency $\Omega_a$ exhibits a minimum, but it is not essential since it is either too small (less than 10 %) or at a relatively small ratio $X_\infty / X_0$, which is physically unrealistic (the distance between the final and initial states of the transferred atom is of the order of $X_0$ corresponding to $X_\infty/X_0 = 2$). More essential is the dependence of the frequency $\Omega_b$, since its decay after the maximum substantially decreases the value of $\Omega_b$. Hence, in the present case $\Omega_b$ is much smaller than $\Omega_a$, and the pre-exponential factor reduces to its transition state theory expression $k_0 \approx \omega_a/2\pi$. For $\omega_E = \Omega_a$ and $X_\infty = 2X_0$, $k_0$ is equal to 28 ps$^{-1}$. The primary catalytic effect of Cu$^+$ ion is to weaken the chemical bond of the NO molecule by charge transfer. The secondary catalytic effect of Cu$^+$ in the bimolecular mechanism is to attract the second NO molecule. In this way the ratio $X_\infty/X_0$ decreases and results in lower activation energy according to Figure 8.

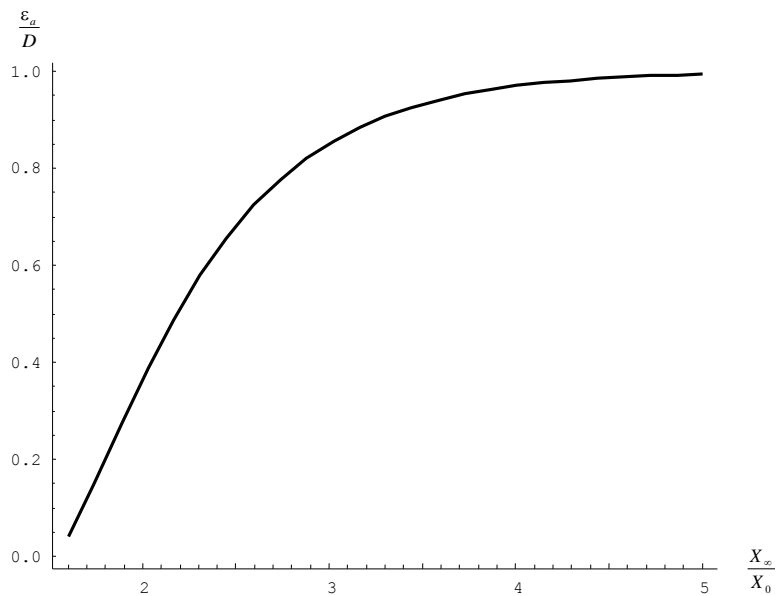

**Fig. 8** Dependence of the activation energy of NO decomposition from the distance between the atoms of the two NO molecules

The present paper relates the catalytic activity of copper-exchanged zeolites on NO decomposition to Cu$^+$ ions. There are two effects of Cu$^+$ on the reaction rate. First, an electron transfer from Cu$^+$ to the NO molecule leads to the weakening of the interaction between the nitrogen and oxygen atoms. Second, a consequent breakdown of the N-O bond under the action of either the zeolite or another NO molecule occurs. In both cases the rate-determining step is the translation of a single atom along with the breaking of the N-O bond. We derived a general formula for the rate constant based on a generalized Langevin equation adopted for the present case. It accounts for both the potential interactions in the system and the memory effects due to the vibrations of the zeolite atoms. The existence of many models of catalytic activity of the Cu-exchanged zeolites on the NO decomposition implies the possibility of multiple reaction paths. Depending on the experimental conditions, one or more of them dominate. For this reason, there are many different mechanisms and catalytic centers reported in the literature. However, because the NO molecule possesses only two atoms one of which is being strongly adsorbed on the copper ion, we believe that the rate-determining step is a single atom transfer in all cases. Hence, our general theory should be applicable regardless of the particular mechanism. Of course, our model possesses a semi-macroscopic character since we did not calculate the exact geometry of the molecular complexes, and the interaction potential $\Phi_0$ has to be known *a priori*. We believe, however, that the developed theory combined with first-principle calculations of the equilibrium configuration of the adsorbed species and their interactions will be able to theoretically predict the NO decomposition rate constant.

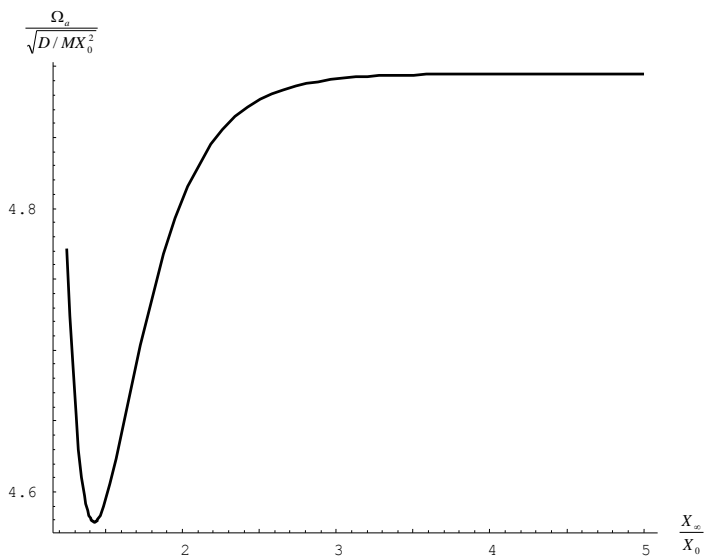

**Fig. 9** Dependence of $\Omega_a$ on the distance between the atoms of the two NO molecules

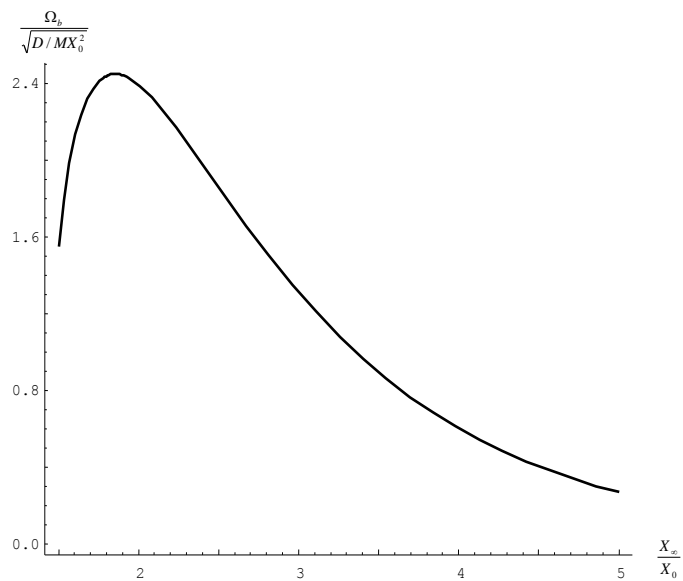

**Fig. 10** Dependence of $\Omega_b$ on the distance between the atoms of the two NO molecules